\documentstyle[osa,twocolumn,epsfig,pstricks]{revtex}

\begin{document}

\title{Generation of correlated photons in controlled
spatial modes by down-conversion in nonlinear waveguides}

\author{Konrad Banaszek, Alfred B. U'Ren, and Ian A. Walmsley}

\address{The Institute of Optics, University of Rochester,
Rochester, New York, 14627}

\date{\today}

\maketitle

\begin{abstract}
We report the observation of correlated photon pairs generated
by spontaneous parametric down-conversion in a quasi-phase matched
KTiOPO$_4$ nonlinear waveguide. The highest ratio of coincidence to
single photon count rates observed in the 830~nm wavelength region exceeds
18\%. This makes nonlinear waveguides a promising source of correlated
photons for metrology and quantum information processing applications.
We also discuss possibilities of controlling the spatial characteristics
of the down-converted photons produced in multimode waveguide structures.
\end{abstract}


\medskip

\noindent
Spontaneous parametric down-conversion is a well-established and
practical method for generating pairs of correlated photons. Such
pairs are a key ingredient of many experiments testing the foundations of
quantum mechanics and in quantum information processing.\cite{ThePhysicsofQI}
Correlated photons have been also used in a number of metrologic
applications.\cite{RariRidlApO97,KwiaSteiPRA93,MigdDatlApO98}

In this paper we report observation of two-photon correlations in the
spontaneous parametric down-conversion generated in a quasi-phase-matched
KTiOPO$_4$ (KTP) waveguide using a $\chi^{(2)}$ nonlinearity.  As a
source of correlated photon pairs, such nonlinear waveguides present
several essential advantages over bulk crystals, the typical medium for
such sources. First, the technique of quasi-phase-matching (QPM) allows
one to utilize larger nonlinear coefficients in some materials, thus
leading to substantially higher two-photon production rates. Secondly,
the waveguide structure provides a means to control precisely the spatial
characteristics of generated photons, as the down-conversion process
is confined to well defined transverse modes.\cite{AndeBeckOpL95}
Appropriate control of the spatio-temporal mode structure
of the photons is a critical
issue in experiments involving interference between multiple photon
pairs.\cite{ThePhysicsofQI} When bulk crystals are used to realize
the parametric process, high visibility of the interference can be
achieved only by heavy spatial filtering of the downconversion signal,
which dramatically reduces the useful fraction of photon pairs. In
contrast, down-conversion in a nonlinear waveguide can provide output in
a single spatial mode, which ensures good interference without filtering
with a significantly larger sample of photon pairs. This feature,
combined with the potential of engineering the temporal properties of
the down-converted photons in QPM structures,\cite{Alfred} opens up
novel possibilities to control the spatio-temporal characteristics of
correlated photon pairs. The use of nonlinear waveguides thus provides
a route to the efficient, if random, generation of ``heralded'' single
photons, alternatively to solid-state sources.\cite{PhotonsOnDemand}

Observation of parametric down-conversion in nonlinear
waveguides has been the subject of three recent experiments. Two of
them\cite{BonfPrunAPL99,TanzDeRiElL01} consisted in detecting
Hanbury Brown-Twiss type coincidences
on the whole down-conversion signal divided by a 50:50 beam splitter. The
third experiment\cite{SanaKawa} measured a Franson-type two-photon
interference effect using an unbalanced Michelson interferometer. A
novel feature of our experiment is that we have undertaken the effort
to separate {\em all} the photon pairs of interest into two different
spatial paths with the help of
a spectrographic setup. Using this technique, we have been
able to demonstrate explicitly strong correlations between photons of
different frequencies. The central objective of our approach was to show
that treating one path as a trigger, one is able to collect effectively
all the conjugate photons in the second path. Thus the quantity
of primary interest in our experiment is the ratio of coincidences to
single trigger counts, and the setup presented below can be viewed as
a scheme for generating single photons in the temporal slots defined by
the pump pulse, with the arrival time known to a femtosecond precision.

\begin{figure}

\psset{unit=0.3in}
\begin{pspicture}(10,11)
\rput[lb](0,0){\epsfig{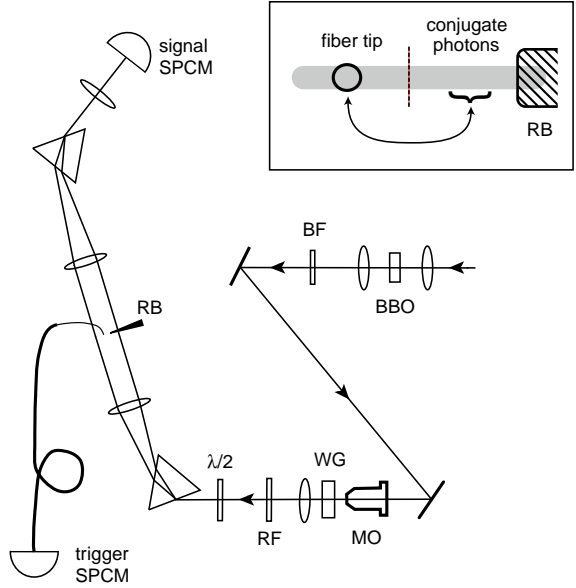}}
\rput[lc](9,8.7){\psline[linewidth=1pt,fillstyle=vlines,%
hatchsep=2pt,linearc=3pt](.7,0)(0,0)(0,1)(.7,1)}
\end{pspicture}

\caption{Experimental setup for detecting correlated pairs of photons.
BBO, beta-barium borate crystal for second harmonic generation; BF, blue
filter; MO, microscope objective; WG, nonlinear waveguide; RF, red filter;
RB, razor blade; SPCM, single photon counting module.
The inset depicts the separation of the signal and trigger photons
performed in the Fourier plane after the first prism.\label{Fig:Setup}}
\end{figure}

The experimental setup is depicted in Fig.~\ref{Fig:Setup}. The output
of a modelocked Ti:Sapphire oscillator is first doubled in a type-I BBO
crystal to generate blue pulses with a central wavelength 418~nm
and a bandwidth of 5~nm, polarized
perpendicularly to the plane of the
figure.
The blue
light is focused using a $20\times$ microscope objective
on the input face of a 1 mm long
quasi-phase-matched KTP waveguide. The production
and the characteristics of the sample used in our experiment
have been described elsewhere.\cite{Bierlein,RoelSunaJAP94}
The light power injected into the
waveguide, measured before the objective, is about 22~$\mu$W. 
The bandwidth
of the down-converted light for the parameters of our experiment
exceeds well over 100~nm, and the down-converted photons have
the same polarization as the pump field.

The
output from the waveguide is coupled out with a laser-diode collimating
lens, and transmitted through an RG665 red filter and a half-wave
plate rotating the polarization by $90^o$.
The separation of the twin photons is performed using a zero dispersion
line employing two Brewster-angle SF10 prisms and two $f=10$~cm lenses.
In the Fourier plane after the first prism,
a multimode fiber tip mounted on a translation stage is used
to collect the trigger photons and send them to a fiber-coupled photon
counting module. The FWHM wavelength range
collected by the fiber tip is about 6~nm.  A second element placed in the
Fourier plane is a razor blade used to block the low wavelength signal
photons which are not conjugated to the trigger photons.  This lowers
the single count rate on the signal detector, thus reducing the number
of accidental coincidences. The remaining signal photons are recombined
into a single beam using a second lens and a
prism, and focused onto
the active area of a free space photon counting module EG\&G
SPCM-AQ-CD2749. The
half-wave plate and all the lenses placed in the down-conversion beam
have a broadband antireflection coating in the 800~nm region.
The position of the fiber tip is calibrated in terms
of the coupled trigger photon wavelengths by sending through the waveguide
light from the laser operated in the cw mode at several
wavelengths in the range 840--870~nm.

The electronic signals from the photon counting modules are shaped using
discriminators to standard NIM logic pulses with a 5~ns width. These
pulses are sent to counters measuring the single rates $R_s$ and $R_t$
of the signal and trigger photons, and also feed inputs of an AND gate,
providing the coincidences whose rate $R_c$ is measured by a third
counter. In Table~\ref{Tab:Results} we present the
count rates obtained over a 300~s counting interval for several positions
of the fiber tip collecting the trigger photons.
The dark count rates, measured with the blocked blue pump beam, were
below 100~Hz for the trigger fiber-coupled module, and below 8~kHz
for the free-space signal module. 

The maximum ratio of coincidence to single counts observed in our
experiment is 18.5\%. 
After correcting for accidentals which results in a minor
few percent change,
this figure can be considered as the overall
detection efficiency of the signal photons, including all the losses since
their generation in the waveguide. The main source of imperfect detection
is probably the non-unit efficiency of the free-space photon counting
module used in the setup, which was optimized for 630~nm, including
the wavelength-specific antireflection coating of the photodiode.
Another important source of losses may originate from the design of the
waveguide itself, which includes non-guiding sections having the same
optical characteristics as the surrounding bulk crystal.\cite{Bierlein}

The highest ratio of coincidences to singles, exceeding 75\% at 702~nm
after corrections for losses, has been so far reported by
Kwiat {\em et al.}\cite{KwiaSteiPRA93} Typical
coincidence count rates observed by them
were roughly ten to twenty times lower than
our data, at similar pump powers.
In their experiment good spatial correlations
between the down-converted photons were ensured by spatial
filtering to a well-defined transverse wave vector, and also
by the use of a long
nonlinear crystal, which introduced more stringent phase-matching
conditions. It should be kept in mind, however, that at the same time
this strengthens the frequency correlations between the down-converted
photons.\cite{GricWalmPRA97}
Consequently, spectral filtering of the trigger photon path
leads to a much narrower bandwidth of the the signal photons.
The essential advantage of nonlinear waveguides is that the shorter
length allows for an efficient generation of ``heralded'' photons
in broadband, possibly femtosecond, wavepackets, while retaining
good spatial control.
We note that the generation of
single photons in well defined spatio-temporal modes is possible also
in thin bulk crystals by appropriate filtering of trigger photons,
as demonstrated in a recent experiment by Lvovsky
{\em et al.},\cite{Lvovsky} but at the cost of extremely low production
rates.

The nonlinear waveguide used in our experiment supports several modes
at both the pump and down-conversion wavelengths.\cite{RoelSunaJAP94}
However, we will now show the multimode
structure of the down-conversion signal can be in principle suppressed
by exploiting the modal dispersion in the waveguide.
In order to discuss this, let us consider
the wave function describing the generated photon
pairs.\cite{GricWalmPRA97}
In a waveguide, this wave function
is a superposition of the terms of the following form:
\begin{eqnarray}
\label{Eq:WaveFunction}
| \psi_{lmn} \rangle  & = &  \int d\omega_s \int d\omega_i \,
\alpha(\omega_s + \omega_i)
\Phi_{lmn}(\omega_s, \omega_i)
\nonumber \\ & & \times
\hat{a}^\dagger_{m} (\omega_s) \hat{a}^\dagger_{n} (\omega_i)
|\text{vac}\rangle
,
\end{eqnarray}
where $\omega_s$ and $\omega_i$ define the signal and the idler
frequencies, respectively, $\alpha(\omega)$ is the spectral envelope
of the pump pulse, $\Phi_{lmn}(\omega_s, \omega_i)$ is the phase
matching function, and $\hat{a}^\dagger_{m} (\omega_s)$ and
$\hat{a}^\dagger_{n} (\omega_i)$ are the creation operators
for the specified frequencies and modes.
The indices $lmn$ label the triplet of the pump mode
$l$ and the pair of the signal mode $m$ and idler mode $n$. The relative
weights of the components $|\psi_{lmn}\rangle$
in the complete superposition are defined by the coupling strength
of the pump field to a specific waveguide mode, and by the overlap of
the pump mode with the product of the down-conversion modes.

\begin{figure}

\centerline{\epsfig{file=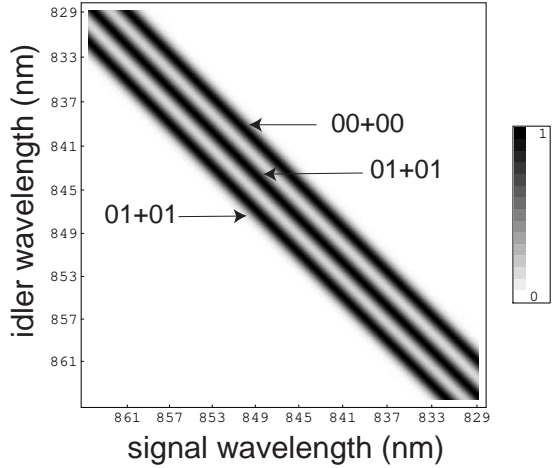,width=3in}}

\caption{Phase matching functions for down-conversion from the $l=00$
pump mode to several pairs of the signal and idler modes.
The first-order QPM
structure is assumed to have the period 3.9375~$\mu$m and the length
1.3~mm. All the parameters of the waveguide are taken from
Ref.~\protect\onlinecite{RoelSunaJAP94}.\label{Fig:PhaseMatching}
For convenience, the frequency
axes are labeled with the corresponding wavelengths.}
\end{figure}

In Fig.~\ref{Fig:PhaseMatching} we plot
$\Phi_{lmn}(\omega_s,\omega_i)$ for several different triplets of the
modes, using the dispersion data from Ref.~\onlinecite{RoelSunaJAP94}. It
is seen that the phase matching functions corresponding to different
triplets occupy separate regions of the plane of the frequencies
$\omega_s$ and $\omega_i$. According to Eq.~(\ref{Eq:WaveFunction}),
down-converted photons are generated only for pairs of frequencies
$\omega_s$ and $\omega_i$ whose sum lies within the bandwidth of
the pump pulse. Thus, if we select the pump pulse spectral amplitude
$\alpha(\omega_s + \omega_i)$ such that it overlaps only with a single
phase matching function $\Phi_{lmn}(\omega_s,\omega_i)$, then the signal
and idler photons are effectively generated in single spatial modes.

In conclusion, we have observed correlated photon pairs in controlled
spatial modes generated by down-conversion in a nonlinear waveguide.
Compared to down-conversion in bulk crystals,
this process has several important advantages, including high
brightness and control of the spatial characteristics of the produced
photons. These features make nonlinear waveguides a promising source
of nonclassical radiation in quantum information and metrology applications.

We thank M. G. Raymer, C. Radzewicz, L. Mandel, and N. Bigelow for
loans of equipment, and acknowledge support from National Science
Foundation and ARO--administered MURI grant No.\ DAAG-19-99-1-0125.

\clearpage

\begin{table}
\caption{The signal $R_s$, trigger $R_t$, coincidence $R_c$, and
accidental coincidence $R_{\text{acc}}$ count
rates for several positions of the fiber tip, labeled
with the central wavelength $\lambda_c$ coupled to the fiber. The
statistical errors are of the order of or smaller
than the lowest digits shown.\label{Tab:Results}}
\begin{tabular}{rrrrrr}
\hline
\multicolumn{1}{c}{$\lambda_c$} &
\multicolumn{1}{c}{$R_s$} &
\multicolumn{1}{c}{$R_t$} &
\multicolumn{1}{c}{$R_c$} & 
\multicolumn{1}{c}{$R_c/R_t$} \\
\multicolumn{1}{c}{(nm)} &
\multicolumn{1}{c}{(kHz)} &
\multicolumn{1}{c}{(Hz)} &
\multicolumn{1}{c}{(Hz)} & 
\multicolumn{1}{c}{(\%)} \\
\hline
909 & 726 & 3755 & 671  & 17.87 \\
897 & 582 & 4866 & 859  & 17.66 \\
885 & 702 & 5692 & 1055 & 18.54 \\
872 & 584 & 6397 & 1171 & 18.31 \\
860 & 403 & 7473 & 1341 & 17.94 \\
848 & 277 & 8149 & 1409 & 17.29 \\
\hline
\end{tabular}
\end{table}

\end{document}